\documentclass[12pt,aps,preprint,nofootinbib,superscriptaddress,nobalancelastpage]{revtex4}

\usepackage{hyperref}
\usepackage{epsfig}
\usepackage{amssymb}
\usepackage{amsmath}
\usepackage{amsfonts}
\usepackage{graphics}
\usepackage{cancel}
\usepackage{bbm}
\usepackage{mathrsfs}
\usepackage{color}

\newcommand{\Ref}{Ref.}
\newcommand{\Sec}{Sec.}

\newcommand{\Tab}{Tab.}

\newcommand{\eq}{Eq.}

\newcommand{\fig}{Fig.}

\newcommand{\bld}[1]{\boldsymbol{#1}}
\newcommand{\bea}{\begin{eqnarray}}
\newcommand{\eea}{\end{eqnarray}}

\newcommand{\be}{\begin{equation}}
\newcommand{\ee}{\end{equation}}

\newcommand{\ba}{\begin{array}}
\newcommand{\ea}{\end{array}}

\newcommand{\ie}{\emph{i.e.}}
\newcommand{\eg}{\emph{e.g.}}

\allowdisplaybreaks

\begin{document}
\title{Aidnogenesis via Leptogenesis and Dark Sphalerons}

\author{Mattias Blennow}
\email[]{blennow@mppmu.mpg.de}
\affiliation{Max-Planck-Institut f\"ur Physik
(Werner-Heisenberg-Institut), F\"ohringer Ring 6, 80805 M\"unchen,
Germany}

\author{Basudeb Dasgupta}
\email[]{dasgupta.10@osu.edu}
\affiliation{CCAPP, The Ohio State University, 191 W.\ Woodruff Avenue, Columbus 43210 USA}

\author{\mbox{Enrique~Fernandez-Martinez}}
\email[]{enfmarti@mppmu.mpg.de}
\affiliation{Max-Planck-Institut f\"ur Physik
(Werner-Heisenberg-Institut), F\"ohringer Ring 6, 80805 M\"unchen,
Germany}

\author{Nuria Rius}
\email[]{nuria@ific.uv.es}
\affiliation{Departamento de F\'isica Te\'orica, IFIC, Universidad de Valencia-CSIC
Edificio de Institutos de Paterna, Apt.\ 22085, 46071 Valencia, Spain}

\begin{abstract}

We discuss aidnogenesis\footnote{From ancient Greek $\alpha\iota\delta\nu{}o\varsigma$ (dark, unseen) and $\gamma\epsilon\nu\epsilon\sigma\iota\varsigma$ (generation, origin).\\}, i.e. the generation of a dark matter asymmetry, via new sphaleron 
processes associated to an extra non-abelian gauge symmetry 
common to both the visible and the dark sectors.
Such a theory can naturally produce an abundance of asymmetric dark matter 
which is of the same size as the lepton and baryon asymmetries, as suggested 
by the similar sizes of the observed baryonic and dark matter energy content, 
and provide a definite prediction for the mass of the dark matter particle. 
We discuss in detail a minimal
realization in which the Standard Model is only extended by dark matter 
fermions which form ``dark baryons'' through an $SU(3)$ interaction, and a 
(broken) horizontal symmetry that induces the new sphalerons. 
The dark matter mass is predicted to be $\sim 6$~GeV, close to the region 
favored by DAMA and CoGeNT. Furthermore, a remnant of the horizontal symmetry 
should be broken at a lower scale and can also explain the Tevatron dimuon anomaly.

\end{abstract}


\preprint{MPP-2010-125, IFIC/10-32, FTUV-10-0909}

\maketitle

\section{Introduction} The building blocks of the Universe are reasonably well-known from a cosmological point of view~\cite{Komatsu:2010fb}. In particular, we now know that about 73~\% of the energy density is in the form of dark energy and causes the accelerated expansion of the Universe while the remaining 27~\% is composed of matter. Baryonic matter makes up only 5~\%, while about five times as much is in the form of a non-luminous weakly interacting species, dubbed ``Dark Matter'' (DM). While this cosmological book-keeping is well developed, the particle nature of DM continues to be one of the most important open questions of particle physics~\cite{Bertone:2004pz}.

In this context, much attention has been devoted to crafting well-motivated and viable theories of DM. The most popular candidates for DM are weakly interacting massive particles~(WIMPs). WIMPs arise naturally in theories, such as supersymmetry, which provide a 
solution to the hierarchy problem and include a ``natural'' DM candidate~\cite{Jungman:1995df} once a discrete symmetry, \eg, $R$-parity, is introduced, so that the least massive particle charged under the new symmetry is stable or very long lived~\cite{Barbier:2004ez}.
This kind of DM comes with the WIMP miracle, \ie, the correct interaction cross-section to thermally produce the density of DM in the early Universe~\cite{Kolb:1990vq}. In hindsight, this becomes the most appealing feature of such DM. The same is true for extra-dimensional models where KK parity ensures the stability of the DM candidates~\cite{Servant:2002aq,Cheng:2002ej}.
 In this scenario, the closeness of the dark matter and baryonic energy densities is merely 
 a coincidence, since they are produced by unrelated mechanisms. However, their similarity suggests that they originated from the same source.
This is the case in models of asymmetric dark matter (ADM), a relatively old idea \cite{Nussinov:1985xr,Barr:1990ca,Barr:1991qn,Kaplan:1991ah} which has been exploited in several models for DM generation \cite{Kuzmin:1996he,Kusenko:1998yi,Farrar:2004qy,Hooper:2004dc,Kitano:2004sv,Agashe:2004bm,Kitano:2005ge,Cosme:2005sb,Farrar:2005zd,Suematsu:2005kp,Tytgat:2006wy,Banks:2006xr,Page:2007sh,Kitano:2008tk,Nardi:2008ix} and has recently received a rising interest \cite{Kaplan:2009ag,Kribs:2009fy,Cohen:2009fz,Cai:2009ia,An:2009vq,Frandsen:2010yj,An:2010kc,Cohen:2010kn,Taoso:2010tg,Shelton:2010ta,Davoudiasl:2010am,Haba:2010bm,Belyaev:2010kp,Chun:2010hz,Buckley:2010ui,Gu:2010ft}. In ADM models, the DM is made up of charge-neutral Dirac fermions, just like baryonic matter and unlike the SUSY neutralino which is a Majorana fermion. Thus, the DM we see today is not generated thermally in the early Universe, but through a particle-antiparticle asymmetry in its production mechanism tied to the production of ordinary matter, giving rise to similar number densities for ordinary matter and DM. Therefore, models of ADM often predict DM masses of $\mathcal O(1)$~GeV, except in instances where the ADM number density can be Boltzmann suppressed or tuned to result in a larger mass~(see, \eg, \Ref~\cite{Nardi:2008ix,Kribs:2009fy,Chun:2010hz}). The phenomenology of ADM is therefore quite different from that of a thermal DM relic. In particular, the prospects of indirect detection of DM in these 
models are suppressed, since the DM does not annihilate 
if only an asymmetric component is present. Nevertheless, there could still be some indirect effects, such as the effect of accreting ADM in the core of stars~\cite{Frandsen:2010yj,Taoso:2010tg}.

As was stressed in \Ref~\cite{Kaplan:1991ah}, any process that creates a baryon or lepton asymmetry could be intimately tied to the creation of ADM as well. In most ADM models, this production of a DM asymmetry is seeded by the observed baryon asymmetry and transferred to the dark sector. However, there has been a recent interest in the opposite mechanism, namely, the creation of a DM asymmetry that is subsequently converted partially into the SM~\cite{Shelton:2010ta,Davoudiasl:2010am,Haba:2010bm,Buckley:2010ui}. 
In this work, we instead assume that the asymmetries in both baryons and DM are created simultaneously by the same processes, as suggested by their similar abundances. First, a lepton asymmetry from the decay of heavy right-handed neutrinos, as in leptogenesis \cite{Fukugita:1986hr}, is induced and then this lepton number is partially converted into both baryon and DM numbers through new sphaleron processes. 
In order to achieve this, we will assume that there is an additional 
non-abelian gauge symmetry group $G$, 
under which both the DM sector and the
SM fermions are charged.
For definiteness, we will consider an additional $SU(2)$ symmetry, but other 
symmetry groups could also fulfill our purposes. 
This extra gauge symmetry could arise from some unified theory at higher 
scales, although this is not required. 
Apart from the right-handed neutrinos and the extra gauge symmetry 
(spontaneously broken by scalar SM singlets),  
we only need to introduce  
new fermion fields $X$, that are singlets under the 
SM gauge group but couple to the new gauge symmetry, and will  
provide the DM candidates.

We include a QCD-like gauge interaction for the DM fermions that prevents their mixing with neutrinos, ensuring their stability without any ad hoc discrete symmetry. The similarity with the SM QCD interaction is also suggestive, given the similar masses of baryons and DM required to fit the observed energy densities. As the DM is essentially composed of dark baryons, it scatters with itself through the QCD-like gauge interaction. This self-interaction can be quite large and leads to almost spherical DM halos in galaxies, in somewhat better agreement with data, compared to WIMPs~\cite{Spergel:1999mh}. The current best limit on the interaction strength comes from observed ellipticities of DM halos of galaxies~\cite{Taoso:2007qk}. However, limits from colliding galaxy clusters, though slightly weaker, are thought to be more robust. These observations tell us that the QCD-like gauge interaction cannot be much stronger than that between baryons.

\section{Aidnogenesis}
As stated in the introduction, a general feature of ADM models is that the production mechanisms for the DM are somehow related to the baryon asymmetry in order to obtain similar abundances. For example, early proposals~\cite{Barr:1990ca,Barr:1991qn,Kaplan:1991ah} include charging light DM fermions under $SU(2)_L$ in order to fix its abundance together with the baryon and lepton asymmetries as a result of the thermal equilibrium in the early Universe~\cite{Khlebnikov:1988sr,Harvey:1990qw}. If dark and baryonic matter have similar number densities, present cosmology data favour rather light
values of the DM fermions mass. Thus, collider constraints rule out the possibility of DM being 
part of a (pure) $SU(2)_L$ multiplet. We instead propose that the DM abundance can be the result of sphaleron 
processes due to a new gauge symmetry that connects the DM and SM sectors.
In such a setting, we extend the SM gauge group by an additional non-abelian gauge symmetry $G$ under which the DM is charged. In order to generate a DM number $X$, related to additional fermions introduced in the theory, we also need to charge some of the SM fermions under this gauge group, generally leading to the associated sphalerons violating the global charges such that $\Delta X = \Delta B/k_1 = \Delta L/k_2$, where $k_i$ are constants. Since the extra sphalerons are responsible for the creation of the DM number, we will refer to them as ``dark sphalerons''  and to the process of the conversion of lepton or baryon number into DM number as ``aidnogenesis''. 
Together with the SM sphalerons, which satisfy $\Delta B = \Delta L$ and $\Delta X = 0$, this will result in a single conserved quantity $(k_2-k_1)X+B-L$, when both sphalerons are active. This quantity is what needs to be generated in the early Universe in order to accommodate both the baryon and DM numbers and is the equivalent to $B-L$ in models of leptogenesis. In the model given below, we will assume that leptogenesis produces an initial $L$ number, which is then partially transformed into $B$ and $X$ by sphaleron processes, we therefore have baryogenesis and aidnogenesis via leptogenesis.

To compute the abundance of $B$ and $X$, we must first write down the equilibrium equations for the chemical potentials of different particle species when both SM and dark sphalerons are active. These equations will be dependent on the specific extension of the SM, but generally lead to a relation of the form
\begin{equation}
 \frac{X}{B-L} = r,
\end{equation}
where $r$ is an $\mathcal O(1)$ constant. Once the additional sphalerons become inactive, the SM sphalerons will continue to act in the standard way, leading to the equilibrium relation $B = 28(B-L)/79$ (in the case of three fermion generations and one Higgs doublet in the standard sector), while conserving $X$ and $B-L$. Thus, for the final relation between the baryon and DM numbers, we obtain
\begin{equation}
 \frac{X}{B} = \frac{79}{28} r.
\end{equation}
This results in a prediction for the DM mass through
\begin{eqnarray}
 \frac{\Omega_B}{\Omega_X} &=& \frac{\rho_B}{\rho_X} = \frac{m_p}{m_{DM}} \left|\frac{B}{X}\right| \simeq 0.2 \nonumber \\
  \Rightarrow 
 &m_{DM} &\simeq 5 m_p \frac{28}{79|r|} \simeq 1.77 \frac{m_p}{|r|}.
\end{eqnarray}
Thus, with $r$ being $\mathcal O(1)$, we would expect the DM mass to be close to the proton mass. It should be noted that the DM mass is here the mass of the lightest of the additional fermions which is stable, since DM number is conserved at the Lagrangian level.

\section{An illustrative model}

{}From the previous section, the question of whether or not aidnogenesis can be achieved in a specific model realization naturally follows. In this section, we will give an example of such realization in a concrete model and compute the corresponding DM density and properties.

For this specific realization, we extend the SM gauge group with an additional $SU(2)_H \times SU(3)_{DC}$, where the $SU(2)_H$ is a horizontal symmetry introduced to provide the dark sphalerons. We have chosen an $SU(2)_H$ for the simplicity of the discussion but, in principle, other non-abelian gauge symmetries encompassing all three fermion generations could be considered as well to address the flavor puzzle. The additional dark color (DC) group $SU(3)_{DC}$ is a color-like gauge interaction in the dark sector.
The DM candidate in this model is a charge neutral $SU(3)_{DC}$ baryon. The fermionic field content in the model is given in \Tab~\ref{tab:ModI} with the corresponding charges. Notice that the choice of which fermion generations form the $SU(2)_H$ doublets is arbitrary and only affects the constraints that can be set with present data on flavour changing neutral current (FCNC) processes on the scale at which the symmetry is broken, but not the generation of the baryon and DM asymmetry. In \Tab~\ref{tab:ModI} 
the $SU(2)_H$ doublets are composed from the fermions of the first two generations, for which the present bounds are strongest, but other possibilities will be briefly discussed in \Sec~\ref{sec:symcomp}. Notice that the only fermion singlet is the right-handed neutrino, for which a Majorana mass is allowed and will be assumed in order to have a see-saw mechanism \cite{Minkowski:1977sc,Gell-Mann:1979ss,Mohapatra:1979ia} for neutrino masses and a lepton number asymmetry in their decay that can seed baryo- and aidnogenesis. This is also consistent with having an even number of $SU(2)_H$ doublets, as required by the cancellation of the Witten anomaly~\cite{Witten:1982fp} for the $SU(2)$ groups. For the same reason, the left-handed components of the DM fermions should not form $SU(2)_L$ doublets. Moreover, since the DM fermions are Dirac, we can define a global DM number $X$, that will be conserved at the Lagrangian level just as the baryon number $B$. 

We further assume that the scalar sector is such that it can provide the required  mass terms for the fermions, either directly or through higher-dimensional operators, after the breaking of both, 
the electroweak and the $SU(2)_H$ symmetries. 
For instance, a minimal realization consists of an extra $SU(2)_H$ doublet, besides the SM Higgs, 
with Yukawa couplings given in terms of effective $d=5$ operators such as
\begin{equation}
O_{d=5} =  c_\alpha \overline{L_H} \Phi_H \Phi_L^\dagger L_{L\alpha},
\end{equation}
where $\Phi_{H,L}$ is the $SU(2)_{H,L}$ Higgs.
After breaking of the $SU(2)_H$ symmetry, this reproduces the SM Yukawa terms.
Just like the $d=5$ Weinberg operator for neutrino masses, such operators can be generated  in several different ways.
The relevant phenomenological aspects  of the model are independent
of the scalar sector, so we shall not discuss it in more detail.

\begin{table}
\begin{center}
\begin{tabular}{|c|c|c|c|c|c|}
\hline
{\bf Field} & $\bld{Y}$ & $\bld{L}$ & $\bld{H}$ & $\bld{C}$ & $\bld{{DC}}$\\
\hline
\hline
$L_{L\alpha}$ ($\nu_{\alpha L}$, $\ell_{\alpha L}$) & $-1/2$ & $\bld 2$ & 1 & 1 & 1 \\
\hline
$L_H$ ($e_R$, $\mu_R$) & $-1$ & 1 & $\bld 2$ & 1 & 1 \\
\hline
$\tau_R$ & $-1$ & 1 & 1 & 1 & 1 \\
\hline
$\nu_{\alpha R}$ & 0 & 1 & 1 & 1 & 1 \\
\hline
$Q_{\alpha L}$ ($u_{\alpha L}$, $d_{\alpha L}$) & $1/6$ & $\bld 2$ & 1 & $\bld 3$ & 1 \\
\hline
$Q_H^u$ ($u_R$, $c_R$) & $2/3$ & 1 & $\bld 2$ & $\bld 3$ & 1 \\
\hline
$Q_H^d$ ($d_R$, $s_R$) & $-1/3$ & 1 & $\bld 2$ & $\bld 3$ & 1 \\
\hline
$t_R$ & $2/3$ & 1 & 1 & $\bld 3$ & 1 \\
\hline
$b_R$ & $-1/3$ & 1 & 1 & $\bld 3$ & 1 \\
\hline
$X_H$ ($x^1_R$, $x^2_R$) & 0 & 1 & $\bld 2$ & 1 & $\bld 3$ \\
\hline
$x^3_R$, $x^\alpha_L$ & 0 & 1 & 1 & 1 & $\bld 3$ \\
\hline
\end{tabular}
\end{center}
\caption{Fermion field content for our illustrative model and the corresponding charge assignments. Whenever a field has an index $\alpha$, the model contains three copies of this field. Note that the assignment of putting particular generations in the $SU(2)_H$ doublets is arbitrary.}
\label{tab:ModI}
\end{table}

In this model, which is free of gauge anomalies, dark sphalerons satisfy $\Delta B/2 = \Delta X = \Delta L$, while the SM sphalerons satisfy $\Delta B = \Delta L$ as usual, resulting in an overall conservation of $B-X-L$ at scales where both sphalerons are active and conservation of both $B-L$ and $X$ separately in the intermediate regime, where only the dark sphalerons are turned off. Thus, if an initial $L$ asymmetry is produced in the decay of the heavy Majorana right-handed neutrinos, both kinds of sphalerons will try to erase it. However, since $B-X-L$ is exactly conserved by the combination of both sphalerons, net $B$ and $X$ asymmetries will be induced. In order to obtain the precise ratios between the final $B$, $L$ and $X$ asymmetries the equilibrium equations for the chemical potentials have to be written down. We find that
\begin{equation}
 r = -\frac{22}{79}
\end{equation}
when both sphaleron processes are in equilibrium. Incidentally, since the DM fields are not charged under any of the SM groups, the equilibrium equations for these groups are identical before and after the $SU(2)_H$ freeze-out. For the final DM to baryon ratio, we obtain
\begin{equation}
 \frac{X}{B} \ \longrightarrow \ - \frac{11}{14}.
\end{equation}
Therefore, in order to accommodate the observed values of $\Omega_X$ and $\Omega_B$, we must have
\begin{equation}
 m_{DM} \simeq m_B \frac{14\Omega_X}{11\Omega_B} = 5.94 \pm 0.42~{\rm GeV},
\label{eq:dm-mass}
\end{equation}
including the errors of the WMAP7 measurementes \cite{Komatsu:2010fb}.
This value is in the low mass regime between 5 and 10~GeV favored by the claimed DM signals of the DAMA/LIBRA~\cite{Bernabei:2010mq} and CoGeNT~\cite{Aalseth:2010vx} collaborations. In particular it is
strikingly close to the $\sim 7$~GeV required to consistently describe both signals and is within the 99~\% confidence level for the mass obtained in \Ref~\cite{Hooper:2010uy}.

In order to obtain the prediction for the DM mass of \eq~(\ref{eq:dm-mass}) we have introduced a scalar doublet of the $SU(2)_H$ symmetry and considered the necessary operators involving this scalar doublet and the Higgs field in order to obtain Yukawa couplings for all fermions. Thus, for the first two generations of charged SM fermions, the Yukawa couplings become effective $d=5$ operators. However, we find that the prediction of \eq~(\ref{eq:dm-mass}) is fairly independent of the particles that mediate the $d=5$ operators as well as the extra $SU(2)_H$ scalar multiplets introduced to break the horizontal symmetry and generate Yukawa couplings. Indeed, if a heavy mediator to induce the effective $d=5$ operator is considered and the effective operator is opened into renormalizable ones,
solving the equilibrium equations for the chemical potential of the heavy particle, the same equation implied by the original $d=5$ operator is obtained. Thus, if the mediator does not carry baryon or DM number (being for instance a scalar) or if it is heavy enough to be integrated out of the theory at the scale at which baryogenesis and aidnogenesis take place, \eq~(\ref{eq:dm-mass}) would not be modified. Regarding the details of the scalar sector introduced to break the $SU(2)_H$ symmetry, we find that the solution of the equilibrium equations imply that the chemical potentials of both the gauge bosons associated to the $SU(2)_H$ symmetry and the scalar multiplets introduced to break it are zero. Thus, the chemical potentials for the particles of the different generations are the same. If different scalar multiplets were assumed instead, setting their chemical potentials to zero would also be a solution to the equilibrium equations and thus \eq~(\ref{eq:dm-mass}) seems
 to be independent also from the details of the $SU(2)_H$ symmetry breaking.

Note that our model has three generations of DM fermions, in analogy with the observed fermions of the SM. However, the third generation of DM fields (\ie, $x^3$) does not play an active role and can be removed from the model. This would result in the DM number density being reduced by a factor $2/3$ and thus predict  $m_{DM} \simeq 9$~GeV. 

\section{Constraints on the model}

\label{sec:symcomp}

The first requisite that has to be met for successful generation of the baryon and dark matter asymmetries via leptogenesis, is that the $SU(2)_H$ sphalerons reach thermal equilibrium before the phase transition occurs and suppresses their rate. Thus, the sphaleron rate should be greater than the Hubble rate, leading to a lower bound on the $SU(2)_H$ coupling constant as a function of the temperature above which equilibrium should be achieved:
\begin{equation}
\alpha_H^4 = \left( \frac{g_H^2}{4 \pi} \right)^4 \gtrsim 10 \frac{T}{M_{Pl}}.
\label{eq:sphalerons}
\end{equation}

On the other hand, a lower bound on the strength of the $SU(2)_H$ interaction at lower energies can be derived from the requirement that the thermally produced symmetric component of dark matter is transfered to the SM fast enough. Indeed, as with any field which is kept in thermal equilibrium, the ADM candidate will have a thermal abundance of both particles and anti-particles in the early Universe. In order for the DM to become asymmetric, there must exist interactions through which this symmetric thermal abundance can be effectively annihilated once the ADM falls out of thermal equilibrium. In our example model, this is achieved by the strong $SU(3)_{DC}$ interactions connecting the symmetric part of the DM fields into dark $SU(3)_{DC}$ mesons which can decay to SM particles via $SU(2)_H$ gauge bosons. Analogously to pion decays in the SM, the dark meson decays will require a chirality flip of the SM fermions they decay into. To estimate the decay rate of the dark mesons we will assume that the dominant decay channel is either to two muons or a tau and a lighter lepton and thus proportional to the muon or tau mass. 

\begin{figure}[t!]
\vspace{-0.5cm}
\begin{center}
\begin{tabular}{cc}
\hspace{-0.55cm} \epsfxsize7.5cm\epsffile{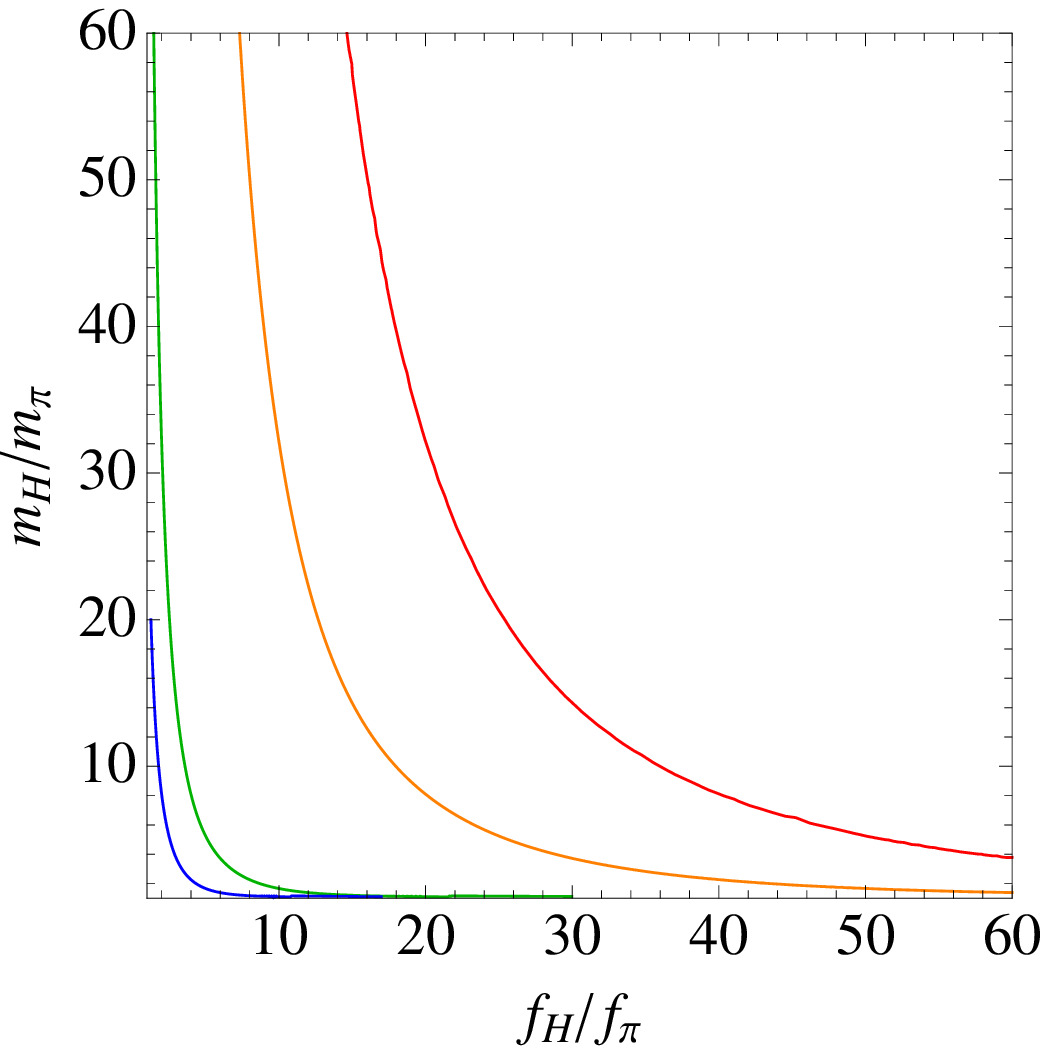} & 
                 \epsfxsize7.5cm\epsffile{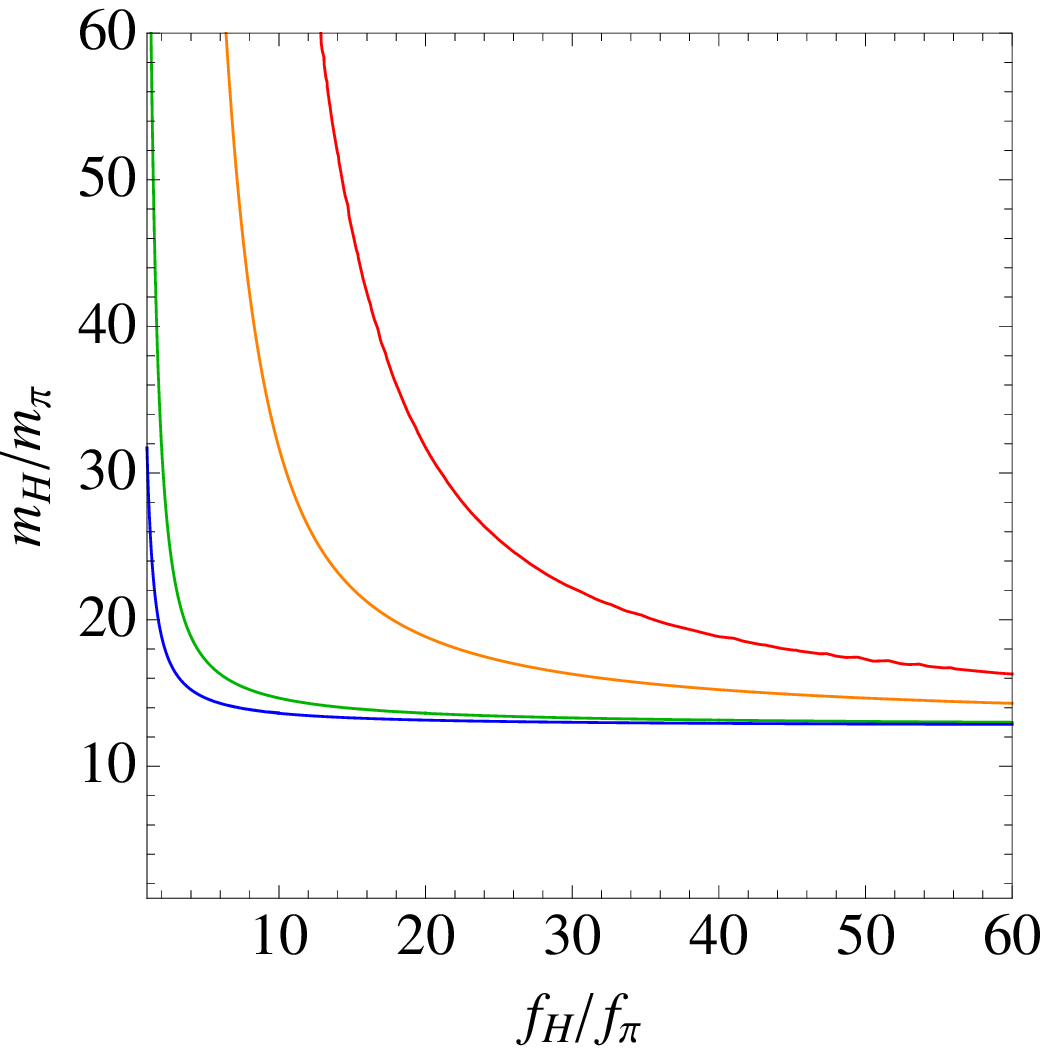} 
\end{tabular}
\caption{\label{fig:bounds}
Contours for the lower bound on $G_F^H$ such that the lifetime of the dark mesons is smaller than $10^{-2}$~s as a function of the dark meson mass $m_H$ and decay constant $f_H$. Left panel for a dominant decay into muons depicts the contours for $G_F^H > 10^{-9}$, $5 \cdot 10^{-10}$, $10^{-10}$~GeV$^{-2}$ and $5 \cdot 10^{-11}$ from the bottom of the plot to the top. Right panel for a dominant decay into a tau and a lighter lepton depicts the contours for $G_F^H > 10^{-10}$, $5 \cdot 10^{-11}$, $10^{-11}$ and $5 \cdot 10^{-12}$~GeV$^{-2}$ from the bottom of the plot to the top. }
\end{center}
\end{figure}

In order not to disturb the standard history of the Universe, the $SU(3)_{DC}$ mesons (which constitute a large matter component) must decay sufficiently fast into SM fermions so that they are no longer present during big bang nucleosynthesis (BBN). Thus, the lifetime of the mesons must be significantly less than one second. Since the horizontal gauge bosons could also induce FCNC processes in 
the SM sector, bounds can be derived
on the related effective $SU(2)_H$ Fermi constant $G_F^H = \sqrt{2}g_H^2/8M_H^2$. Naturally, such bounds are stronger if the two SM generations involved contain the lightest fermions of each type (\eg, $e$ and $\mu$ rather than $\mu$ and $\tau$), since these FCNC have stronger experimental constraints. In \fig~\ref{fig:bounds} we show contours for the lower bound on $G_F^H$ such that the lifetime of the dark mesons is smaller than $10^{-2}$~s as a function of the dark meson mass $m_H$ and decay constant $f_H$. Notice that the values of these two quantities will depend on the strength of the $SU(3)_{DC}$ interaction as well as the masses of the dark matter fermions. In any case the ``dark meson'' masses should be heavier than $\sim 100$ MeV so that they decay before BBN but lighter than the ``dark baryon'' mass so that the symmetric component is stored mainly in mesons and not baryons and antibaryons, this corresponds to the maximum value of $m_H$ depicted in \fig~\ref{fig:bounds}.
If the dominant decay channel is to a muon pair (left panel), then the bound is typically of $\mathcal O(10^{-10})$~GeV$^{-2}$, while the bounds are about an order of magnitude weaker for decays into a tau and a lighter lepton (right panel). In this last case the decay can only happen if $m_H > m_\tau$ which corresponds to the horizontal asymptote.

On the other hand, the strongest constraint on $G_F^H$ from FCNC stems from the bound on the decay $K \to e \mu$ and implies that $G_F^H < 3.6\cdot 10^{-12}$~GeV$^{-2}$, which would cause tension with the lower bounds derived in \fig~\ref{fig:bounds}.
However, this constraint does not apply if the horizontal symmetry is broken in stages. For example, this may be achieved by first breaking the $SU(2)_H$ to $U(1)_H$ by a real scalar triplet acquiring a vacuum expectation value along the $\sigma_3$ direction and giving large masses to the flavor changing gauge bosons while leaving the flavor conserving one massless. This procedure is similar to Georgi and Glashow's model of electroweak interactions, which did not include neutral currents~\cite{Georgi:1972cj}.
The remaining flavor diagonal $U(1)_H$ can be subsequently broken at a scale low enough to obtain interactions of the strength required for the dark mesons to decay into two muons. In this case the dominant decay channel of the dark mesons could be to two muons and the constraints on the left-handed panel of \fig~\ref{fig:bounds} would apply. Reconciling this scenario with the lower bound on $g_H$ from \eq~(\ref{eq:sphalerons}) is easy since the scale of the $SU(2)_H$ symmetry breaking is unrelated to the mass of the flavour-conserving $Z'$, for instance with $g_H = 0.5$ the sphalerons would reach thermal equilibrium at $T \lesssim 10^{11}$ GeV and the first stage of symmetry breaking, which would freeze out the sphalerons, should occur above $\Lambda \gtrsim 10^5$ GeV. 

An interesting alternative would be to couple the two heaviest SM fermion generations to the $SU(2)_H$ and having the scalar triplet acquire its vev along the $\sigma_1$ direction. In this case the residual $U(1)_H$ would still induce FCNC in the $t-c$, $b-s$ and $\tau-\mu$ sectors but the constraints are in this case weaker and allow for sufficiently fast dark meson decays. Indeed, such a $U(1)_H$ with strength $G_F^H=7 \cdot 10^{-11}$~GeV$^{-2}$ is allowed by present data and would contribute to CP violation in the $B_s$ system, accommodating the observed dimuon anomaly at Tevatron~\cite{Park:2010sg}. In this case the dark mesons would decay into a tau and a muon and the constraints depicted in the right panel of \fig~\ref{fig:bounds} would apply. In this scenario, it is also possible to identify the scale of the $SU(2)_H$ symmetry breaking with the mass of the flavour-changing $Z'$ contributing to $B_s$ mixing. In combination with 
\eq~(\ref{eq:sphalerons}) this would imply 
that $g_H \gtrsim  0.06$ and $\Lambda \gtrsim 2.7$ TeV, for the
dark sphalerons to enter in thermal equilibrium
at $T \gtrsim \Lambda$.

\section{Phenomenological prospects}
\label{sec:pheno}
The detection prospects for this kind of DM mainly come from three directions: direct detection through nuclear recoil, production at colliders, and through FCNCs. 
The most interesting prospects are from direct detection by observing the nuclear recoil of a DM particle hitting a nucleon. 
Given that the cross-section goes as $G_{F}^{H2} E^2$, where $E$ is an energy at the typical scale of the DM mass in our model, we predict a ADM-nucleon cross-section $\sigma \gtrsim  10^{-46}$~cm$^2$, where we have used the dark meson lifetime constraint $G_{F}^{H} \gtrsim 10^{-10}$~GeV$^{-2}$. It is worth pointing out that generally the sensitivity of direct detection experiments is weaker for the low DM mass regime favored by ADM. In this regime, the present bounds set on the spin-independent cross section by the XENON100 \cite{Aprile:2010um} and CDMS \cite{Ahmed:2009zw} collaborations are $\mathcal O(10^{-40})$~cm$^2$. However, proposed upgrades to these experiments or future experiments could probe most of the allowed range.

From the collider perspective, the best 
present bounds come from the non-observation of the $Z'$ associated to the $U(1)_H$. The results from LEP-II suggest that, if the $Z'$ at low energies couples with full strength to electrons and other charged leptons then $G_{F}^{H} < 5.14\cdot10^{-8}$~GeV$^{-2}$~\cite{:2003ih}, while the present limits from Tevatron are subdominant. This would be in tension with the interpretation of the DAMA and CoGeNT experiments that require~$G_{F}^{H} \sim 10^{-7}$GeV$^{-2}$~\cite{Hooper:2010uy}. These bounds could be evaded if the $SU(2)_H$ gauge symmetry couples mainly to the heavier charged lepton generations. In any case, the bound is about two or three orders of magnitude above the dark meson lifetime constraint. 
There is scope to improve this at LHC, where the limits can be improved by almost 30~\% in some regions of the $g_{Z'}/M_{Z'}$ plane. On the other hand, future LHC constraints could be derived from a somewhat general approach, as shown in \Ref~\cite{Goodman:2010ku}, where one could explore cross-sections well-below $10^{-41}$~cm$^2$ and almost all the way to $10^{-46}$~cm$^2$, depending on the details of the coupling of DM to quarks.

Furthermore, FCNC processes are generic features of this type of model. However, the interaction strength for these depends on the specifics of symmetry breaking and can be either strong enough to be already making an impact at present experiments (\ie, the Tevatron dimuon anomaly) or so weak that it is unobservable for all practical purposes (\ie, the symmetry is broken in such a way that only a flavor diagonal interaction remains at lower energies). If FCNCs are discovered at Super-B factories~\cite{O'Leary:2010af}, the flavor symmetry breaking structure could be probed.

\section{Summary and conclusions}

In this work, we have discussed the prospects of generating asymmetric dark matter (ADM) from an initial lepton asymmetry through the sphalerons of a new non-abelian gauge group. The initial lepton asymmetry is taken to be the result of a leptogenesis mechanism. As an example of a model where this occurs, we have discussed an anomaly free model, in which the Standard Model (SM) is extended by dark matter fermion fields and an additional horizontal $SU(2)$ gauge symmetry. In addition, the model contains a color like $SU(3)$ gauge interaction in the dark sector, preventing the mixing with the right-handed neutrinos and requiring the dark matter to be in the form of charge neutral baryon-like states. 
The mass of these baryon-like states is predicted by comparing the number densities of baryons and dark baryons with the ratio $\Omega_{X}/\Omega_{B} = 4.98 \pm 0.35$ and the result is $m_{DM} = (5.94 \pm 0.42)$~GeV.
The similarity with the SM QCD interaction is also suggestive, given the similar masses of baryons and DM required to fit the observed energy densities. The $SU(3)$ self-interaction can be quite large and leads to almost spherical DM halos in galaxies, in somewhat better agreement with data, compared to WIMPs.

In order to get rid of the thermally produced symmetric component of the dark matter before big bang nucleosynthesis (BBN), it is required that the dark mesons decay into SM fields with a coupling constant of $\gtrsim 10^{-10}$~GeV$^{-2}$. Thus, the predictions of this model are suggestively close to the values 
favoured by a consistent description of the DAMA and CoGeNT experiments, 
which require $m_{DM} \sim 7$~GeV and a DM-baryon interaction with 
strength~$\sim G_F/100$.

In the particular realization we discussed, the additional chiral symmetry corresponding to the dark sphalerons is a horizontal symmetry inspired by possible solutions to the flavor puzzle and the Tevatron dimuon anomaly. It is beyond the scope of this work to build a full model addressing these two problems in combination with dark matter generation. However, the possibility of a connection between dark matter and the flavor puzzle is intriguing and should be explored further.

\section{Acknowledgments}
The authors would like to acknowledge useful discussions with S.~Antusch, J.~Beacom, C.~Biggio, B.~Gavela, G.~Raffelt, J.~Redondo, C.~Rott, F.~Steffen and S.~Zhou. We are especially indebted to L.~Lopez Honorez for her participation in the early stages of the project. B.D.\ and E.F.M.\ would like to acknowledge support from MPI for Physics in Munich, where this work began, and for hospitality in the later stages of this work. M.B.\ , E.F.M.\  and N.R.\ would also like to acknowledge 
the hospitality of CERN, especially during the $\nu$TheME workshop, where 
part of this work was performed. This work was supported by
the European Union through the European Commission Marie Curie Actions Framework Programme 7 Intra-European Fellowship: Neutrino Evolution [M.B.], 
by Spanish  MICCIN  under grants FPA-2007-60323
and Consolider-Ingenio PAU (CSD2007-00060) and 
by Generalitat Valenciana grant PROMETEO/2009/116 [N.R.].


\begin{thebibliography}{61}
\expandafter\ifx\csname natexlab\endcsname\relax\def\natexlab#1{#1}\fi
\expandafter\ifx\csname bibnamefont\endcsname\relax
  \def\bibnamefont#1{#1}\fi
\expandafter\ifx\csname bibfnamefont\endcsname\relax
  \def\bibfnamefont#1{#1}\fi
\expandafter\ifx\csname citenamefont\endcsname\relax
  \def\citenamefont#1{#1}\fi
\expandafter\ifx\csname url\endcsname\relax
  \def\url#1{\texttt{#1}}\fi
\expandafter\ifx\csname urlprefix\endcsname\relax\def\urlprefix{URL }\fi
\providecommand{\bibinfo}[2]{#2}
\providecommand{\eprint}[2][]{\url{#2}}

\bibitem[{\citenamefont{Komatsu et~al.}(2010)}]{Komatsu:2010fb}
\bibinfo{author}{\bibfnamefont{E.}~\bibnamefont{Komatsu}} \bibnamefont{et~al.}
  (\bibinfo{year}{2010}), \eprint{arXiv:1001.4538}.

\bibitem[{\citenamefont{Bertone et~al.}(2005)\citenamefont{Bertone, Hooper, and
  Silk}}]{Bertone:2004pz}
\bibinfo{author}{\bibfnamefont{G.}~\bibnamefont{Bertone}},
  \bibinfo{author}{\bibfnamefont{D.}~\bibnamefont{Hooper}}, \bibnamefont{and}
  \bibinfo{author}{\bibfnamefont{J.}~\bibnamefont{Silk}},
  \bibinfo{journal}{Phys. Rept.} \textbf{\bibinfo{volume}{405}},
  \bibinfo{pages}{279} (\bibinfo{year}{2005}), \eprint{hep-ph/0404175}.

\bibitem[{\citenamefont{Jungman et~al.}(1996)\citenamefont{Jungman,
  Kamionkowski, and Griest}}]{Jungman:1995df}
\bibinfo{author}{\bibfnamefont{G.}~\bibnamefont{Jungman}},
  \bibinfo{author}{\bibfnamefont{M.}~\bibnamefont{Kamionkowski}},
  \bibnamefont{and} \bibinfo{author}{\bibfnamefont{K.}~\bibnamefont{Griest}},
  \bibinfo{journal}{Phys. Rept.} \textbf{\bibinfo{volume}{267}},
  \bibinfo{pages}{195} (\bibinfo{year}{1996}), \eprint{hep-ph/9506380}.

\bibitem[{\citenamefont{Barbier et~al.}(2005)}]{Barbier:2004ez}
\bibinfo{author}{\bibfnamefont{R.}~\bibnamefont{Barbier}} \bibnamefont{et~al.},
  \bibinfo{journal}{Phys. Rept.} \textbf{\bibinfo{volume}{420}},
  \bibinfo{pages}{1} (\bibinfo{year}{2005}), \eprint{hep-ph/0406039}.

\bibitem[{\citenamefont{Kolb and Turner}(1990)}]{Kolb:1990vq}
\bibinfo{author}{\bibfnamefont{E.~W.} \bibnamefont{Kolb}} \bibnamefont{and}
  \bibinfo{author}{\bibfnamefont{M.~S.} \bibnamefont{Turner}},
  \bibinfo{journal}{Front.Phys.} \textbf{\bibinfo{volume}{69}},
  \bibinfo{pages}{1} (\bibinfo{year}{1990}).

\bibitem[{\citenamefont{Servant and Tait}(2003)}]{Servant:2002aq}
\bibinfo{author}{\bibfnamefont{G.}~\bibnamefont{Servant}} \bibnamefont{and}
  \bibinfo{author}{\bibfnamefont{T.~M.~P.} \bibnamefont{Tait}},
  \bibinfo{journal}{Nucl. Phys.} \textbf{\bibinfo{volume}{B650}},
  \bibinfo{pages}{391} (\bibinfo{year}{2003}), \eprint{hep-ph/0206071}.

\bibitem[{\citenamefont{Cheng et~al.}(2002)\citenamefont{Cheng, Feng, and
  Matchev}}]{Cheng:2002ej}
\bibinfo{author}{\bibfnamefont{H.-C.} \bibnamefont{Cheng}},
  \bibinfo{author}{\bibfnamefont{J.~L.} \bibnamefont{Feng}}, \bibnamefont{and}
  \bibinfo{author}{\bibfnamefont{K.~T.} \bibnamefont{Matchev}},
  \bibinfo{journal}{Phys. Rev. Lett.} \textbf{\bibinfo{volume}{89}},
  \bibinfo{pages}{211301} (\bibinfo{year}{2002}), \eprint{hep-ph/0207125}.

\bibitem[{\citenamefont{Nussinov}(1985)}]{Nussinov:1985xr}
\bibinfo{author}{\bibfnamefont{S.}~\bibnamefont{Nussinov}},
  \bibinfo{journal}{Phys.Lett.} \textbf{\bibinfo{volume}{B165}},
  \bibinfo{pages}{55} (\bibinfo{year}{1985}).

\bibitem[{\citenamefont{Barr et~al.}(1990)\citenamefont{Barr, Chivukula, and
  Farhi}}]{Barr:1990ca}
\bibinfo{author}{\bibfnamefont{S.~M.} \bibnamefont{Barr}},
  \bibinfo{author}{\bibfnamefont{R.~S.} \bibnamefont{Chivukula}},
  \bibnamefont{and} \bibinfo{author}{\bibfnamefont{E.}~\bibnamefont{Farhi}},
  \bibinfo{journal}{Phys. Lett.} \textbf{\bibinfo{volume}{B241}},
  \bibinfo{pages}{387} (\bibinfo{year}{1990}).

\bibitem[{\citenamefont{Barr}(1991)}]{Barr:1991qn}
\bibinfo{author}{\bibfnamefont{S.~M.} \bibnamefont{Barr}},
  \bibinfo{journal}{Phys. Rev.} \textbf{\bibinfo{volume}{D44}},
  \bibinfo{pages}{3062} (\bibinfo{year}{1991}).

\bibitem[{\citenamefont{Kaplan}(1992)}]{Kaplan:1991ah}
\bibinfo{author}{\bibfnamefont{D.~B.} \bibnamefont{Kaplan}},
  \bibinfo{journal}{Phys. Rev. Lett.} \textbf{\bibinfo{volume}{68}},
  \bibinfo{pages}{741} (\bibinfo{year}{1992}).

\bibitem[{\citenamefont{Kuzmin}(1998)}]{Kuzmin:1996he}
\bibinfo{author}{\bibfnamefont{V.~A.} \bibnamefont{Kuzmin}},
  \bibinfo{journal}{Phys. Part. Nucl.} \textbf{\bibinfo{volume}{29}},
  \bibinfo{pages}{257} (\bibinfo{year}{1998}), \eprint{hep-ph/9701269}.

\bibitem[{\citenamefont{Kusenko}(1998)}]{Kusenko:1998yi}
\bibinfo{author}{\bibfnamefont{A.}~\bibnamefont{Kusenko}}
  (\bibinfo{year}{1998}), \eprint{hep-ph/9901353}.

\bibitem[{\citenamefont{Farrar and Zaharijas}(2004)}]{Farrar:2004qy}
\bibinfo{author}{\bibfnamefont{G.~R.} \bibnamefont{Farrar}} \bibnamefont{and}
  \bibinfo{author}{\bibfnamefont{G.}~\bibnamefont{Zaharijas}}
  (\bibinfo{year}{2004}), \eprint{hep-ph/0406281}.

\bibitem[{\citenamefont{Hooper et~al.}(2005)\citenamefont{Hooper,
  March-Russell, and West}}]{Hooper:2004dc}
\bibinfo{author}{\bibfnamefont{D.}~\bibnamefont{Hooper}},
  \bibinfo{author}{\bibfnamefont{J.}~\bibnamefont{March-Russell}},
  \bibnamefont{and} \bibinfo{author}{\bibfnamefont{S.~M.} \bibnamefont{West}},
  \bibinfo{journal}{Phys. Lett.} \textbf{\bibinfo{volume}{B605}},
  \bibinfo{pages}{228} (\bibinfo{year}{2005}), \eprint{hep-ph/0410114}.

\bibitem[{\citenamefont{Kitano and Low}(2005{\natexlab{a}})}]{Kitano:2004sv}
\bibinfo{author}{\bibfnamefont{R.}~\bibnamefont{Kitano}} \bibnamefont{and}
  \bibinfo{author}{\bibfnamefont{I.}~\bibnamefont{Low}},
  \bibinfo{journal}{Phys. Rev.} \textbf{\bibinfo{volume}{D71}},
  \bibinfo{pages}{023510} (\bibinfo{year}{2005}{\natexlab{a}}),
  \eprint{hep-ph/0411133}.

\bibitem[{\citenamefont{Agashe and Servant}(2005)}]{Agashe:2004bm}
\bibinfo{author}{\bibfnamefont{K.}~\bibnamefont{Agashe}} \bibnamefont{and}
  \bibinfo{author}{\bibfnamefont{G.}~\bibnamefont{Servant}},
  \bibinfo{journal}{JCAP} \textbf{\bibinfo{volume}{0502}}, \bibinfo{pages}{002}
  (\bibinfo{year}{2005}), \eprint{hep-ph/0411254}.

\bibitem[{\citenamefont{Kitano and Low}(2005{\natexlab{b}})}]{Kitano:2005ge}
\bibinfo{author}{\bibfnamefont{R.}~\bibnamefont{Kitano}} \bibnamefont{and}
  \bibinfo{author}{\bibfnamefont{I.}~\bibnamefont{Low}}
  (\bibinfo{year}{2005}{\natexlab{b}}), \eprint{hep-ph/0503112}.

\bibitem[{\citenamefont{Cosme et~al.}(2005)\citenamefont{Cosme, Lopez~Honorez,
  and Tytgat}}]{Cosme:2005sb}
\bibinfo{author}{\bibfnamefont{N.}~\bibnamefont{Cosme}},
  \bibinfo{author}{\bibfnamefont{L.}~\bibnamefont{Lopez~Honorez}},
  \bibnamefont{and} \bibinfo{author}{\bibfnamefont{M.~H.~G.}
  \bibnamefont{Tytgat}}, \bibinfo{journal}{Phys. Rev.}
  \textbf{\bibinfo{volume}{D72}}, \bibinfo{pages}{043505}
  (\bibinfo{year}{2005}), \eprint{hep-ph/0506320}.

\bibitem[{\citenamefont{Farrar and Zaharijas}(2006)}]{Farrar:2005zd}
\bibinfo{author}{\bibfnamefont{G.~R.} \bibnamefont{Farrar}} \bibnamefont{and}
  \bibinfo{author}{\bibfnamefont{G.}~\bibnamefont{Zaharijas}},
  \bibinfo{journal}{Phys. Rev. Lett.} \textbf{\bibinfo{volume}{96}},
  \bibinfo{pages}{041302} (\bibinfo{year}{2006}), \eprint{hep-ph/0510079}.

\bibitem[{\citenamefont{Suematsu}(2006)}]{Suematsu:2005kp}
\bibinfo{author}{\bibfnamefont{D.}~\bibnamefont{Suematsu}},
  \bibinfo{journal}{Astropart. Phys.} \textbf{\bibinfo{volume}{24}},
  \bibinfo{pages}{511} (\bibinfo{year}{2006}), \eprint{hep-ph/0510251}.

\bibitem[{\citenamefont{Tytgat}(2006)}]{Tytgat:2006wy}
\bibinfo{author}{\bibfnamefont{M.~H.~G.} \bibnamefont{Tytgat}}
  (\bibinfo{year}{2006}), \eprint{hep-ph/0606140}.

\bibitem[{\citenamefont{Banks et~al.}(2006)\citenamefont{Banks, Echols, and
  Jones}}]{Banks:2006xr}
\bibinfo{author}{\bibfnamefont{T.}~\bibnamefont{Banks}},
  \bibinfo{author}{\bibfnamefont{S.}~\bibnamefont{Echols}}, \bibnamefont{and}
  \bibinfo{author}{\bibfnamefont{J.~L.} \bibnamefont{Jones}},
  \bibinfo{journal}{JHEP} \textbf{\bibinfo{volume}{11}}, \bibinfo{pages}{046}
  (\bibinfo{year}{2006}), \eprint{hep-ph/0608104}.

\bibitem[{\citenamefont{Page}(2007)}]{Page:2007sh}
\bibinfo{author}{\bibfnamefont{V.}~\bibnamefont{Page}}, \bibinfo{journal}{JHEP}
  \textbf{\bibinfo{volume}{04}}, \bibinfo{pages}{021} (\bibinfo{year}{2007}),
  \eprint{hep-ph/0701266}.

\bibitem[{\citenamefont{Kitano et~al.}(2008)\citenamefont{Kitano, Murayama, and
  Ratz}}]{Kitano:2008tk}
\bibinfo{author}{\bibfnamefont{R.}~\bibnamefont{Kitano}},
  \bibinfo{author}{\bibfnamefont{H.}~\bibnamefont{Murayama}}, \bibnamefont{and}
  \bibinfo{author}{\bibfnamefont{M.}~\bibnamefont{Ratz}},
  \bibinfo{journal}{Phys. Lett.} \textbf{\bibinfo{volume}{B669}},
  \bibinfo{pages}{145} (\bibinfo{year}{2008}), \eprint{arXiv:0807.4313}.

\bibitem[{\citenamefont{Nardi et~al.}(2009)\citenamefont{Nardi, Sannino, and
  Strumia}}]{Nardi:2008ix}
\bibinfo{author}{\bibfnamefont{E.}~\bibnamefont{Nardi}},
  \bibinfo{author}{\bibfnamefont{F.}~\bibnamefont{Sannino}}, \bibnamefont{and}
  \bibinfo{author}{\bibfnamefont{A.}~\bibnamefont{Strumia}},
  \bibinfo{journal}{JCAP} \textbf{\bibinfo{volume}{0901}}, \bibinfo{pages}{043}
  (\bibinfo{year}{2009}), \eprint{arXiv:0811.4153}.

\bibitem[{\citenamefont{Kaplan et~al.}(2009)\citenamefont{Kaplan, Luty, and
  Zurek}}]{Kaplan:2009ag}
\bibinfo{author}{\bibfnamefont{D.~E.} \bibnamefont{Kaplan}},
  \bibinfo{author}{\bibfnamefont{M.~A.} \bibnamefont{Luty}}, \bibnamefont{and}
  \bibinfo{author}{\bibfnamefont{K.~M.} \bibnamefont{Zurek}},
  \bibinfo{journal}{Phys. Rev.} \textbf{\bibinfo{volume}{D79}},
  \bibinfo{pages}{115016} (\bibinfo{year}{2009}), \eprint{arXiv:0901.4117}.

\bibitem[{\citenamefont{Kribs et~al.}(2010)\citenamefont{Kribs, Roy, Terning,
  and Zurek}}]{Kribs:2009fy}
\bibinfo{author}{\bibfnamefont{G.~D.} \bibnamefont{Kribs}},
  \bibinfo{author}{\bibfnamefont{T.~S.} \bibnamefont{Roy}},
  \bibinfo{author}{\bibfnamefont{J.}~\bibnamefont{Terning}}, \bibnamefont{and}
  \bibinfo{author}{\bibfnamefont{K.~M.} \bibnamefont{Zurek}},
  \bibinfo{journal}{Phys. Rev.} \textbf{\bibinfo{volume}{D81}},
  \bibinfo{pages}{095001} (\bibinfo{year}{2010}), \eprint{arXiv:0909.2034}.

\bibitem[{\citenamefont{Cohen and Zurek}(2010)}]{Cohen:2009fz}
\bibinfo{author}{\bibfnamefont{T.}~\bibnamefont{Cohen}} \bibnamefont{and}
  \bibinfo{author}{\bibfnamefont{K.~M.} \bibnamefont{Zurek}},
  \bibinfo{journal}{Phys. Rev. Lett.} \textbf{\bibinfo{volume}{104}},
  \bibinfo{pages}{101301} (\bibinfo{year}{2010}), \eprint{arXiv:0909.2035}.

\bibitem[{\citenamefont{Cai et~al.}(2009)\citenamefont{Cai, Luty, and
  Kaplan}}]{Cai:2009ia}
\bibinfo{author}{\bibfnamefont{Y.}~\bibnamefont{Cai}},
  \bibinfo{author}{\bibfnamefont{M.~A.} \bibnamefont{Luty}}, \bibnamefont{and}
  \bibinfo{author}{\bibfnamefont{D.~E.} \bibnamefont{Kaplan}}
  (\bibinfo{year}{2009}), \eprint{arXiv:0909.5499}.

\bibitem[{\citenamefont{An et~al.}(2010{\natexlab{a}})\citenamefont{An, Chen,
  Mohapatra, and Zhang}}]{An:2009vq}
\bibinfo{author}{\bibfnamefont{H.}~\bibnamefont{An}},
  \bibinfo{author}{\bibfnamefont{S.-L.} \bibnamefont{Chen}},
  \bibinfo{author}{\bibfnamefont{R.~N.} \bibnamefont{Mohapatra}},
  \bibnamefont{and} \bibinfo{author}{\bibfnamefont{Y.}~\bibnamefont{Zhang}},
  \bibinfo{journal}{JHEP} \textbf{\bibinfo{volume}{03}}, \bibinfo{pages}{124}
  (\bibinfo{year}{2010}{\natexlab{a}}), \eprint{arXiv:0911.4463}.

\bibitem[{\citenamefont{Frandsen and Sarkar}(2010)}]{Frandsen:2010yj}
\bibinfo{author}{\bibfnamefont{M.~T.} \bibnamefont{Frandsen}} \bibnamefont{and}
  \bibinfo{author}{\bibfnamefont{S.}~\bibnamefont{Sarkar}},
  \bibinfo{journal}{Phys. Rev. Lett.} \textbf{\bibinfo{volume}{105}},
  \bibinfo{pages}{011301} (\bibinfo{year}{2010}), \eprint{arXiv:1003.4505}.

\bibitem[{\citenamefont{An et~al.}(2010{\natexlab{b}})}]{An:2010kc}
\bibinfo{author}{\bibfnamefont{H.}~\bibnamefont{An}} \bibnamefont{et~al.},
  \bibinfo{journal}{Phys. Rev.} \textbf{\bibinfo{volume}{D82}},
  \bibinfo{pages}{023533} (\bibinfo{year}{2010}{\natexlab{b}}),
  \eprint{arXiv:1004.3296}.

\bibitem[{\citenamefont{Cohen et~al.}(2010)\citenamefont{Cohen, Phalen, Pierce,
  and Zurek}}]{Cohen:2010kn}
\bibinfo{author}{\bibfnamefont{T.}~\bibnamefont{Cohen}},
  \bibinfo{author}{\bibfnamefont{D.~J.} \bibnamefont{Phalen}},
  \bibinfo{author}{\bibfnamefont{A.}~\bibnamefont{Pierce}}, \bibnamefont{and}
  \bibinfo{author}{\bibfnamefont{K.~M.} \bibnamefont{Zurek}},
  \bibinfo{journal}{Phys. Rev.} \textbf{\bibinfo{volume}{D82}},
  \bibinfo{pages}{056001} (\bibinfo{year}{2010}), \eprint{arXiv:1005.1655}.

\bibitem[{\citenamefont{Taoso et~al.}(2010)}]{Taoso:2010tg}
\bibinfo{author}{\bibfnamefont{M.}~\bibnamefont{Taoso}} \bibnamefont{et~al.}
  (\bibinfo{year}{2010}), \eprint{arXiv:1005.5711}.

\bibitem[{\citenamefont{Shelton and Zurek}(2010)}]{Shelton:2010ta}
\bibinfo{author}{\bibfnamefont{J.}~\bibnamefont{Shelton}} \bibnamefont{and}
  \bibinfo{author}{\bibfnamefont{K.~M.} \bibnamefont{Zurek}}
  (\bibinfo{year}{2010}), \eprint{arXiv:1008.1997}.

\bibitem[{\citenamefont{Davoudiasl et~al.}(2010)\citenamefont{Davoudiasl,
  Morrissey, Sigurdson, and Tulin}}]{Davoudiasl:2010am}
\bibinfo{author}{\bibfnamefont{H.}~\bibnamefont{Davoudiasl}},
  \bibinfo{author}{\bibfnamefont{D.~E.} \bibnamefont{Morrissey}},
  \bibinfo{author}{\bibfnamefont{K.}~\bibnamefont{Sigurdson}},
  \bibnamefont{and} \bibinfo{author}{\bibfnamefont{S.}~\bibnamefont{Tulin}}
  (\bibinfo{year}{2010}), \eprint{arXiv:1008.2399}.

\bibitem[{\citenamefont{Haba and Matsumoto}(2010)}]{Haba:2010bm}
\bibinfo{author}{\bibfnamefont{N.}~\bibnamefont{Haba}} \bibnamefont{and}
  \bibinfo{author}{\bibfnamefont{S.}~\bibnamefont{Matsumoto}}
  (\bibinfo{year}{2010}), \eprint{arXiv:1008.2487}.

\bibitem[{\citenamefont{Belyaev et~al.}(2010)\citenamefont{Belyaev, Frandsen,
  Sarkar, and Sannino}}]{Belyaev:2010kp}
\bibinfo{author}{\bibfnamefont{A.}~\bibnamefont{Belyaev}},
  \bibinfo{author}{\bibfnamefont{M.~T.} \bibnamefont{Frandsen}},
  \bibinfo{author}{\bibfnamefont{S.}~\bibnamefont{Sarkar}}, \bibnamefont{and}
  \bibinfo{author}{\bibfnamefont{F.}~\bibnamefont{Sannino}}
  (\bibinfo{year}{2010}), \eprint{arXiv:1007.4839}.

\bibitem[{\citenamefont{Chun}(2010)}]{Chun:2010hz}
\bibinfo{author}{\bibfnamefont{E.~J.} \bibnamefont{Chun}}
  (\bibinfo{year}{2010}), \eprint{arXiv:1009.0983}.

\bibitem[{\citenamefont{Buckley and Randall}(2010)}]{Buckley:2010ui}
\bibinfo{author}{\bibfnamefont{M.~R.} \bibnamefont{Buckley}} \bibnamefont{and}
  \bibinfo{author}{\bibfnamefont{L.}~\bibnamefont{Randall}}
  (\bibinfo{year}{2010}), \eprint{arXiv:1009.0270}.

\bibitem[{\citenamefont{Gu et~al.}(2010)\citenamefont{Gu, Lindner, Sarkar, and
  Zhang}}]{Gu:2010ft}
\bibinfo{author}{\bibfnamefont{P.-H.} \bibnamefont{Gu}},
  \bibinfo{author}{\bibfnamefont{M.}~\bibnamefont{Lindner}},
  \bibinfo{author}{\bibfnamefont{U.}~\bibnamefont{Sarkar}}, \bibnamefont{and}
  \bibinfo{author}{\bibfnamefont{X.}~\bibnamefont{Zhang}}
  (\bibinfo{year}{2010}), \eprint{arXiv:1009.2690}.

\bibitem[{\citenamefont{Fukugita and Yanagida}(1986)}]{Fukugita:1986hr}
\bibinfo{author}{\bibfnamefont{M.}~\bibnamefont{Fukugita}} \bibnamefont{and}
  \bibinfo{author}{\bibfnamefont{T.}~\bibnamefont{Yanagida}},
  \bibinfo{journal}{Phys. Lett.} \textbf{\bibinfo{volume}{B174}},
  \bibinfo{pages}{45} (\bibinfo{year}{1986}).

\bibitem[{\citenamefont{Spergel and Steinhardt}(2000)}]{Spergel:1999mh}
\bibinfo{author}{\bibfnamefont{D.~N.} \bibnamefont{Spergel}} \bibnamefont{and}
  \bibinfo{author}{\bibfnamefont{P.~J.} \bibnamefont{Steinhardt}},
  \bibinfo{journal}{Phys. Rev. Lett.} \textbf{\bibinfo{volume}{84}},
  \bibinfo{pages}{3760} (\bibinfo{year}{2000}), \eprint{astro-ph/9909386}.

\bibitem[{\citenamefont{Taoso et~al.}(2008)\citenamefont{Taoso, Bertone, and
  Masiero}}]{Taoso:2007qk}
\bibinfo{author}{\bibfnamefont{M.}~\bibnamefont{Taoso}},
  \bibinfo{author}{\bibfnamefont{G.}~\bibnamefont{Bertone}}, \bibnamefont{and}
  \bibinfo{author}{\bibfnamefont{A.}~\bibnamefont{Masiero}},
  \bibinfo{journal}{JCAP} \textbf{\bibinfo{volume}{0803}}, \bibinfo{pages}{022}
  (\bibinfo{year}{2008}), \eprint{arXiv:0711.4996}.

\bibitem[{\citenamefont{Khlebnikov and Shaposhnikov}(1988)}]{Khlebnikov:1988sr}
\bibinfo{author}{\bibfnamefont{S.~Y.} \bibnamefont{Khlebnikov}}
  \bibnamefont{and} \bibinfo{author}{\bibfnamefont{M.~E.}
  \bibnamefont{Shaposhnikov}}, \bibinfo{journal}{Nucl. Phys.}
  \textbf{\bibinfo{volume}{B308}}, \bibinfo{pages}{885} (\bibinfo{year}{1988}).

\bibitem[{\citenamefont{Harvey and Turner}(1990)}]{Harvey:1990qw}
\bibinfo{author}{\bibfnamefont{J.~A.} \bibnamefont{Harvey}} \bibnamefont{and}
  \bibinfo{author}{\bibfnamefont{M.~S.} \bibnamefont{Turner}},
  \bibinfo{journal}{Phys. Rev.} \textbf{\bibinfo{volume}{D42}},
  \bibinfo{pages}{3344} (\bibinfo{year}{1990}).

\bibitem[{\citenamefont{Minkowski}(1977)}]{Minkowski:1977sc}
\bibinfo{author}{\bibfnamefont{P.}~\bibnamefont{Minkowski}},
  \bibinfo{journal}{Phys. Lett.} \textbf{\bibinfo{volume}{B67}},
  \bibinfo{pages}{421} (\bibinfo{year}{1977}).

\bibitem[{\citenamefont{Gell-Mann et~al.}(1979)\citenamefont{Gell-Mann, Ramond,
  and Slansky}}]{Gell-Mann:1979ss}
\bibinfo{author}{\bibfnamefont{M.}~\bibnamefont{Gell-Mann}},
  \bibinfo{author}{\bibfnamefont{P.}~\bibnamefont{Ramond}}, \bibnamefont{and}
  \bibinfo{author}{\bibfnamefont{R.}~\bibnamefont{Slansky}}, in
  \emph{\bibinfo{booktitle}{{\it Complex spinors and unified theories} in {\it
  Supergravity, Proceedings of the Workshop, Stony Brook, New York, 1979}}},
  edited by \bibinfo{editor}{\bibfnamefont{P.}~\bibnamefont{van Nieuwenhuizen}}
  \bibnamefont{and} \bibinfo{editor}{\bibfnamefont{D.}~\bibnamefont{Freedman}}
  (\bibinfo{publisher}{North-Holland, Amsterdam}, \bibinfo{year}{1979}).

\bibitem[{\citenamefont{Mohapatra and Senjanovi{\'c}}(1980)}]{Mohapatra:1979ia}
\bibinfo{author}{\bibfnamefont{R.~N.} \bibnamefont{Mohapatra}}
  \bibnamefont{and}
  \bibinfo{author}{\bibfnamefont{G.}~\bibnamefont{Senjanovi{\'c}}},
  \bibinfo{journal}{Phys. Rev. Lett.} \textbf{\bibinfo{volume}{44}},
  \bibinfo{pages}{912} (\bibinfo{year}{1980}).

\bibitem[{\citenamefont{Witten}(1982)}]{Witten:1982fp}
\bibinfo{author}{\bibfnamefont{E.}~\bibnamefont{Witten}},
  \bibinfo{journal}{Phys.Lett.} \textbf{\bibinfo{volume}{B117}},
  \bibinfo{pages}{324} (\bibinfo{year}{1982}).

\bibitem[{\citenamefont{Bernabei et~al.}(2010)}]{Bernabei:2010mq}
\bibinfo{author}{\bibfnamefont{R.}~\bibnamefont{Bernabei}}
  \bibnamefont{et~al.}, \bibinfo{journal}{Eur. Phys. J.}
  \textbf{\bibinfo{volume}{C67}}, \bibinfo{pages}{39} (\bibinfo{year}{2010}),
  \eprint{arXiv:1002.1028}.

\bibitem[{\citenamefont{Aalseth et~al.}(2010)}]{Aalseth:2010vx}
\bibinfo{author}{\bibfnamefont{C.~E.} \bibnamefont{Aalseth}}
  \bibnamefont{et~al.} (\bibinfo{collaboration}{CoGeNT})
  (\bibinfo{year}{2010}), \eprint{arXiv:1002.4703}.

\bibitem[{\citenamefont{Hooper et~al.}(2010)\citenamefont{Hooper, Collar, Hall,
  and McKinsey}}]{Hooper:2010uy}
\bibinfo{author}{\bibfnamefont{D.}~\bibnamefont{Hooper}},
  \bibinfo{author}{\bibfnamefont{J.~I.} \bibnamefont{Collar}},
  \bibinfo{author}{\bibfnamefont{J.}~\bibnamefont{Hall}}, \bibnamefont{and}
  \bibinfo{author}{\bibfnamefont{D.}~\bibnamefont{McKinsey}}
  (\bibinfo{year}{2010}), \eprint{arXiv:1007.1005}.

\bibitem[{\citenamefont{Georgi and Glashow}(1972)}]{Georgi:1972cj}
\bibinfo{author}{\bibfnamefont{H.}~\bibnamefont{Georgi}} \bibnamefont{and}
  \bibinfo{author}{\bibfnamefont{S.~L.} \bibnamefont{Glashow}},
  \bibinfo{journal}{Phys. Rev. Lett.} \textbf{\bibinfo{volume}{28}},
  \bibinfo{pages}{1494} (\bibinfo{year}{1972}).

\bibitem[{\citenamefont{Park et~al.}(2010)\citenamefont{Park, Shu, Wang, and
  Yanagida}}]{Park:2010sg}
\bibinfo{author}{\bibfnamefont{S.~C.} \bibnamefont{Park}},
  \bibinfo{author}{\bibfnamefont{J.}~\bibnamefont{Shu}},
  \bibinfo{author}{\bibfnamefont{K.}~\bibnamefont{Wang}}, \bibnamefont{and}
  \bibinfo{author}{\bibfnamefont{T.~T.} \bibnamefont{Yanagida}}
  (\bibinfo{year}{2010}), \eprint{arXiv:1008.4445}.

\bibitem[{\citenamefont{Aprile et~al.}(2010)}]{Aprile:2010um}
\bibinfo{author}{\bibfnamefont{E.}~\bibnamefont{Aprile}} \bibnamefont{et~al.}
  (\bibinfo{collaboration}{XENON100}) (\bibinfo{year}{2010}),
  \eprint{arXiv:1005.0380}.

\bibitem[{\citenamefont{Ahmed et~al.}(2010)}]{Ahmed:2009zw}
\bibinfo{author}{\bibfnamefont{Z.}~\bibnamefont{Ahmed}} \bibnamefont{et~al.}
  (\bibinfo{collaboration}{The CDMS-II}), \bibinfo{journal}{Science}
  \textbf{\bibinfo{volume}{327}}, \bibinfo{pages}{1619} (\bibinfo{year}{2010}),
  \eprint{arXiv:0912.3592}.

\bibitem[{\citenamefont{{The LEP Collaborations}}(2003)}]{:2003ih}
\bibinfo{author}{\bibnamefont{{The LEP Collaborations}}}
  (\bibinfo{collaboration}{LEP}) (\bibinfo{year}{2003}),
  \eprint{hep-ex/0312023}.

\bibitem[{\citenamefont{Goodman et~al.}(2010)}]{Goodman:2010ku}
\bibinfo{author}{\bibfnamefont{J.}~\bibnamefont{Goodman}} \bibnamefont{et~al.}
  (\bibinfo{year}{2010}), \eprint{arXiv:1008.1783}.

\bibitem[{\citenamefont{O'Leary et~al.}(2010)}]{O'Leary:2010af}
\bibinfo{author}{\bibfnamefont{B.}~\bibnamefont{O'Leary}} \bibnamefont{et~al.}
  (\bibinfo{collaboration}{SuperB}) (\bibinfo{year}{2010}),
  \eprint{arXiv:1008.1541}.

\end{thebibliography}
\end{document}